\documentclass[twocolumn,showpacs,preprintnumbers,amsmath,amssymb]{revtex4}

\usepackage{graphicx} 
\usepackage{dcolumn}
\usepackage{bm}
\usepackage{amsmath}
\usepackage{multirow}
\usepackage{amssymb}

\usepackage{eepic}
\usepackage{amscd}
\usepackage{longtable}
\usepackage{array}
\usepackage{graphicx}
\usepackage{slashbox}
\usepackage{multirow}

\begin{document}

\title{ Exact dynamical state of the exclusive queueing process with
  deterministic hopping }

\author{Chikashi Arita}
\email{chikashi.arita@cea.fr}
\affiliation{
Institut de Physique Th\'{e}orique,
CEA Saclay,
F-91191 Gif-sur-Yvette, France}
\author{Andreas Schadschneider}
\email{as@thp.uni-koeln.de}
\affiliation{
Institut f\"{u}r Theoretische Physik,
Universit\"{a}t zu K\"{o}ln,
D-50937 K\"{o}ln, Germany}


\newcommand\bra{\langle W|}
\newcommand\ket{|V\rangle}


\begin{abstract}
The exclusive queueing process (EQP) has recently been introduced as a
model for the dynamics of queues which takes into account the spatial
structure of the queue. It can be interpreted as a totally asymmetric
exclusion process of varying length. Here we investigate the case of
deterministic bulk hopping $p=1$ which turns out to be one of the rare
cases where exact nontrivial results for the dynamical properties can
be obtained.
Using a time-dependent matrix product form we calculate several dynamical properties,
 e.g.\ the density profile of the system.
\end{abstract}



\maketitle

\begin{figure}\begin{center}
\centering
\includegraphics{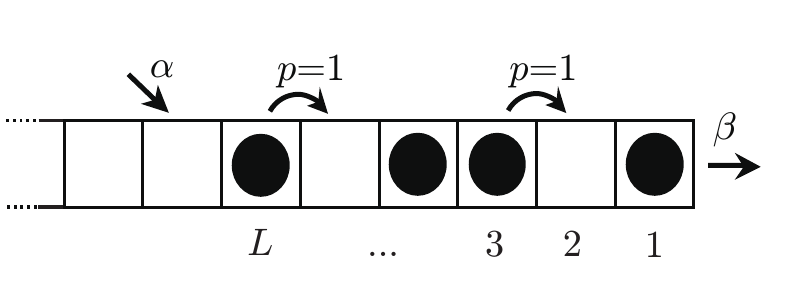}
\caption{Exclusive queueing process (EPQ) with deterministic bulk hopping
($p=1$).}
\label{fig:eq} 
\end{center}\end{figure}

\section{Introduction}\label{sec:intro}

The one-dimensional asymmetric exclusion process, which can be
regarded as the prototypical stochastic interacting particle system
\cite{RefL}, has been intensively studied in view of its
non-equilibrium properties \cite{RefD}, exact solvability
\cite{RefS,RefBE} and applicability to practical problems
\cite{RefSCN}.  The state space for the exclusion process is the set
of configurations of particles (in other words, the exclusion process
has a ``spatial structure''), and each particle can hop to its nearest
neighbor sites only if the target site is empty (``excluded-volume
effect'').

On the other hand, the queueing process
 is one of the basic stochastic
processes in the field of operations research 
\cite{RefMedhi,RefSaaty,Reffactory}.
In addition to its practical relevance 
it often appears as effective model, e.g.\  in all kinds of jamming 
phenomena. Usually the spatial structure of the queue is neglected, 
i.e.\ the queues are regarded as ``compact''.
However, often this assumption is not justified, e.g.\ in pedestrian queues.  
Therefore, recently a queueing process with excluded-volume effect
({\it exclusive queueing process}, EQP) has been proposed
\cite{RefA,RefY,RefAY}.
On a semi-infinite lattice,
particles enter the system at the left site next to the
leftmost occupied site 
and leave the system at the rightmost site.
In the bulk the particles move according to
the rules of the totally asymmetric exclusion process (TASEP), 
see Fig.~\ref{fig:eq}.

The EPQ can be interpreted as a TASEP with varying
system length
which allows to analyze its stationary-state properties 
\cite{RefA,RefY,RefAY}.
In a more recent paper \cite{RefAS},
dynamical properties of the EQP were analyzed.  Especially, for the
deterministic bulk hopping case, 
dynamical behaviors of the average system length and the average
number of particles were investigated exactly.  In this paper we derive more detailed results for the dynamical properties. 

The stationary state of the TASEP and some of its generalizations 
 have been solved by means of the matrix product ansatz in the recent two
decades \cite{RefBE}.
The application of the matrix product ansatz to
the calculation of non-stationary states is quite challenging and has
been achieved only in a few cases so far,
see e.g.\ 
\cite{RefStS1,RefStS2,RefSaWa,RefSchuetz,RefAAMP}.
In this paper we will introduce a matrix product dynamical state for
the EQP, providing an explicit representation for the matrices.  We
utilize it for calculating typical quantities both in queueing
theory and exclusion processes.
  
Here we define the EQP as a discrete-time Markov process on a
semi-infinite chain where sites are labeled by natural numbers from
right to left (Fig.~\ref{fig:eq}).  A new particle enters the chain
with probability $\alpha$ only at the left site next to the leftmost
occupied site ($j=L$).
If there is no particle on the chain, a new particle
enters at the (fixed) rightmost site ($j=1$) with probability $\alpha$.  
Each particle on the chain necessarily hops to its right nearest neighbor
site if it is empty, i.e.\ we consider the limit of deterministic
bulk hopping ($p=1$).
A particle on the rightmost site leaves the
system with probability $\beta$.  These transitions occur
simultaneously within one time step, i.e.\ we apply the
fully parallel update scheme.
Since we restrict our consideration to the 
case of deterministic bulk hopping  $p=1$ 
(the so-called rule 184 cellular automaton),
the stochasticity of the model is due to only the injection and extraction
probabilities $\alpha$ and $\beta$.

\begin{figure}
\begin{center}
\centering
\includegraphics[height=62mm]{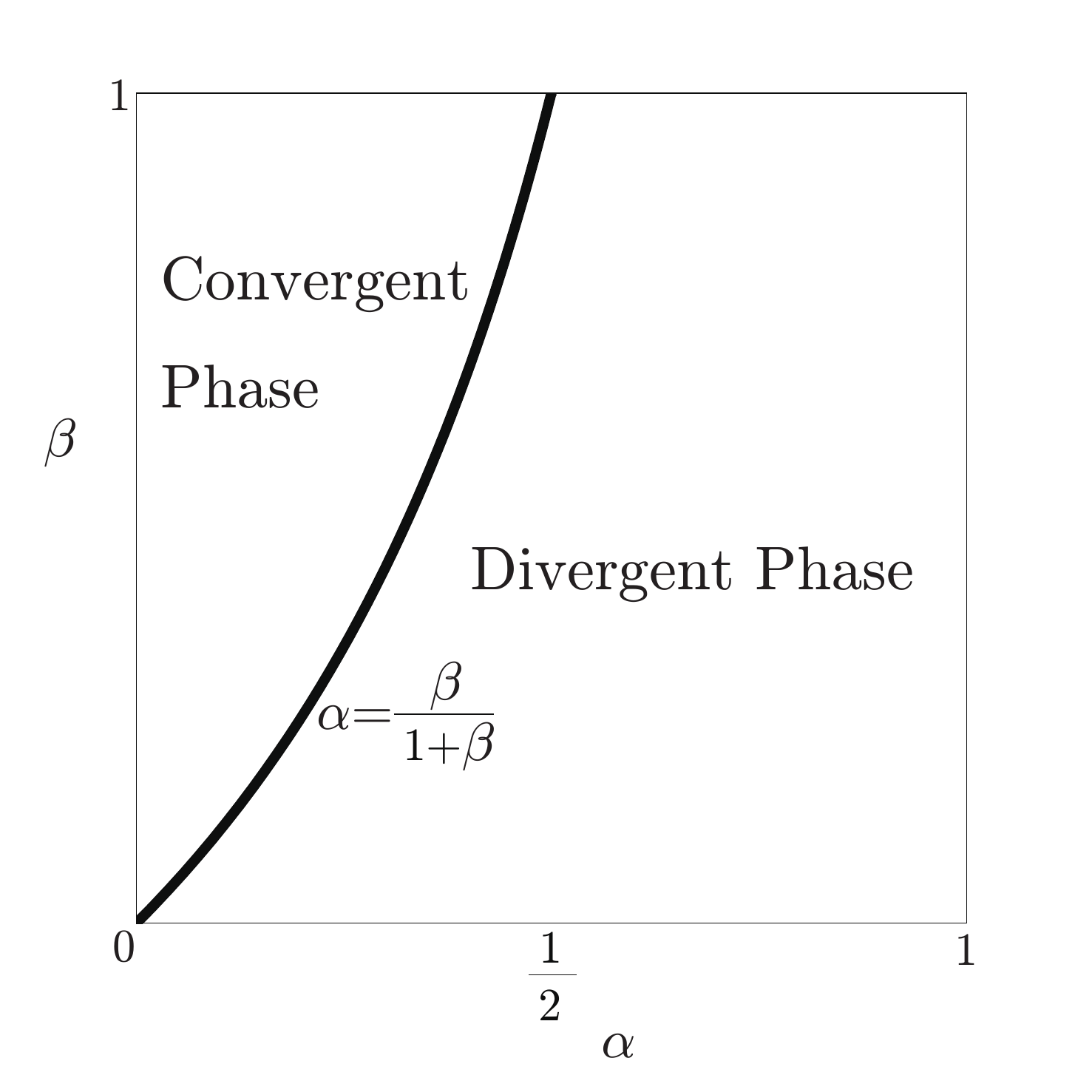}\\[0mm]
\caption{Phase diagram for the EQP with $p=1$.}
\label{fig:pd}
\end{center}\end{figure}

In \cite{RefY,RefAY,RefAS}, the phase diagram of the EQP was derived.
The parameter space is divided
into two regions (Fig.~\ref{fig:pd}),   
the convergent phase $\alpha<\frac{\beta}{1+\beta}$ and the divergent
phase $\alpha>\frac{\beta}{1+\beta}$.  In the convergent phase, the
system approaches a stationary state which can be written in 
a matrix product form.
On the other hand, in the divergent phase, a
stationary state does not exist, and the average length of the system
$\langle L_t\rangle$ and the average number of particles $\langle
N_t\rangle$ increase asymptotically linearly in time $t$.  On the
``critical line'' $\alpha=\frac{\beta}{1+\beta}$, both $\langle
L_t\rangle$ and $\langle N_t\rangle$ exhibit diffusive behavior, 
i.e.\ they increase being asymptotically proportional to $\sqrt{t}$.

In this paper, we investigate the dynamical
(i.e. time-dependent) properties of the EQP in more detail.
In the next section we write down the dynamical state (solution to the
master equation) in a matrix product form. 
Using this form we
investigate the waiting time,
which is one of the basic quantities
in queueing theory, in Sec.~\ref{sec:waiting}.  In
Sec.~\ref{sec:density-current} we determine the density profile and the
particle current profile.
Concluding remarks are given in
Sec.~\ref{sec:conclusion}.
In Appendix we review  results on
the usual (i.e.\ without excluded-volume effect)
 discrete-time queueing process.

\section{Exact dynamical state}\label{sec:mp}

For each site $j$
we define the state variable $\tau_j=1$ or 0 
corresponding to  
being  occupied or unoccupied, respectively.  
For simplicity we impose the initial condition that there is no particle 
in the system, i.e.\ an empty chain.
 The state space is
\begin{align}
 \widetilde{S} 
& = \{ \emptyset \} \cup
    \{ \sigma_\ell \cdots \sigma_1 | \ell \in \mathbb N,
       \sigma_j\in\{1,10\}  \}  \nonumber\\
&=\{ \emptyset, 1,10,11,110,101,1010, 111,  \dots \}  \label{tildeS}
\end{align}
which is a subset $(\widetilde{S} \subset S)$ of
\begin {equation}
S = \{ \emptyset ,1\}
 \cup \{ 1\tau_{L-1} \cdots \tau_1 | L-1 \in \mathbb N,
       \tau_j\in\{1,0\}  \} \, .
\label{S}
\end{equation}
The element $\emptyset$ corresponds to the state where there is no
particle in the system.
Note that, for
$p=1$, the sequence 00 never appears if the system starts from   the empty chain.
For simplification, we do not write the
infinite number of 0's left to the leftmost particle.
We denote the probability of finding a state $\tau\in\widetilde S$ at
time $t$ by $P_t (\tau)$, and the initial condition is written as $
P_0(\emptyset) = 1$ and $ P_0(\tau)=0\ (\tau\in\widetilde S\setminus\{
\emptyset\}) $. We denote the system ``length'' for the state
$\tau\in\widetilde{S}$ or $\tau\in S$ by $|\tau|$, which is nothing
but the position of the leftmost particle (Fig.~\ref{fig:eq}).  In
particular we define $|\emptyset|=0$.  

For the generic choice of the parameters $0<\alpha<1$ and $0<\beta<1$,
the process is irreducible and non-periodic on $\widetilde S$.
The master equation is simply written as
\begin{widetext}
\begin{align}
  \label{master-empty}
 P_{t+1} (\emptyset) &= 
     (1-\alpha)\beta P_t(1) + (1-\alpha)P_t(\emptyset), \\
 P_{t+1} (1) &=  
    (1-\alpha)P_t(10) + (1-\alpha)(1-\beta) P_t(1) + \alpha P_t(\emptyset), \\
  \label{master-u10}
 P_{t+1} (u10) &=  
      (1-\alpha)\beta P_t(1u1) + (1-\alpha)\beta P_t(10u1)
       +  \alpha\beta P_t(u1), \\
 P_{t+1} (u101) &= 
    (1-\alpha)P_t(1u10) + (1-\alpha) P_t(10u10) 
   +  \alpha P_t(u10), \\
 P_{t+1} (u11) &=  (1-\alpha)(1-\beta)P_t(1u1) 
  + (1-\alpha)(1-\beta)P_t(10u1) +  \alpha(1-\beta)P_t(u1)
\label{master-u11}
\end{align}
\end{widetext}
for $u\in \widetilde S$.
In particular, for $u=\emptyset$ we set
$\emptyset 10 = 10$, $1\emptyset 1 =11$,
and so on.  This simple form is due
to the deterministic hopping $p=1$.

We derive an exact dynamical state,
beginning with the factorization ansatz
\begin{align}\label{eq:P=QY}
   P_t(\tau) = Q_t(|\tau|) Y (\tau)\,  .
\end{align}
The first part $Q$ depends only on time
 and the system length, and
the second part $Y$ is independent of time
and satisfies the following relations:
\begin{align}
\label{relation101}
   Y(u_1 101 u_2) =&  \beta Y(u_1 1 u_2 ) ,\\
\label{relation11}
   Y(u_1 11 u_2)   =& (1-\beta) Y(u_1  1  u_2),  \\
\label{relation10}
   Y(u_1 10)   =& \beta Y(u_1  1 ) , \\
\label{relation1}
   Y(1) =& 1, \\
\label{relation0}
   Y(\emptyset) =& 1.
\end{align}
One can easily see that the solution to these relations is
\begin{align}
  Y (\tau_L\cdots \tau_1)
  = \beta^{  \#\{j|\tau_j=0\}  }
  (1-\beta)^{  2\#\{j|\tau_j=1\}  -  L  -  \tau_1  }
\end{align}
for $\tau_L\cdots\tau_1\in\widetilde S \setminus\{\emptyset\}$.
The relations \eqref{relation101}-\eqref{relation1}
also have the following matrix product representation,
which is more convenient later:
\begin{align}
\label{eq:Y=wxxxv}
   Y(\tau_L\cdots \tau_1) 
   = \bra X_{\tau_L}\cdots X_{\tau_1}\ket
\end{align}
with
\begin{align}
  X_1=D= \left(\begin{array}{cc} 1-\beta & 0 \\ \sqrt{\beta} & 0  \end{array}\right),\quad
  X_0=E= \left(\begin{array}{cc} 0  &  \sqrt{\beta} \\0 & 0  \end{array}\right),   \\
 \bra = \left( 1 \  \sqrt{\beta}  \right) , \quad
  \ket = \left(\begin{array}{c} 1 \\  \sqrt{\beta}   \end{array}\right)  .
\end{align}
These are essentially the same matrices and vectors as for the matrix
product \emph{stationary} state for the EQP with $p=1$ \cite{RefAY}.  The
first part $Q_t(L)$ gives the probability that the system length is
$L$ at time $t$ since
\begin{align}
 \sum_{\tau\in \widetilde{S} \atop |\tau|=L} Y (\tau)
 = \sum_{\tau\in S \atop |\tau|=L} Y (\tau)
 =\bra D(D+E)^{L-1} \ket =1   .
\end{align}
Note that we can replace $\widetilde S$ by $S$ in the above equation
thanks to $E^2=0$.  Inserting the relations
\eqref{relation101}-\eqref{relation1} into the master equation
\eqref{master-empty}-\eqref{master-u11}, we obtain
\begin{align}
\label{Q0}
  Q_{t+1}(0) =&  (1-\alpha)Q_t(0) + \beta (1-\alpha)Q_t(1), \\
  Q_{t+1}(L) =&  \alpha Q_t(L-1) + (1-\alpha) (1-\beta)Q_t(L) \nonumber\\
           & +  (1-\alpha) \beta Q_t(L+1) .
\label{eq:QL}
\end{align}
These equations actually agree
 with Eqs.~(65) and (66) in \cite{RefAS} which
were derived in a different way.
The solution to this recurrence
formula with the initial condition
\begin{align}
  Q_0(0)= 1,\quad Q_0(L) =0\ (L\in \mathbb N)
\end{align}
is given by \cite{RefAS}
\begin{align}
\label{eq:Q=}
 Q_t (L) = C_{z^t} 
 \frac{1-\Lambda}{1-z} \Lambda^L 
\end{align}
with
\begin{align}
\label{eq:Lambda=}
   \Lambda &= \frac{1-(1-\alpha)(1-\beta)z - r}{2(1-\alpha)\beta z},  \\
   r &= \sqrt{ [ 1-(1-\alpha)(1-\beta)z ]^2 - 4 (1-\alpha)\alpha\beta z^2 } ,
   \label{eq:r=}
\end{align}
where $C_{z^t}F(z)$ denotes the coefficient of $z^t$
in the Laurent series for the function $F(z)$,
that is $ C_{z^t}F(z) = \oint \frac{dz}{2\pi i z^{t+1} }F(z)$
with a small anti-clockwise path enclosing the origin
of the complex plane.
The average system length  $\langle L_t\rangle $
at time $t$ is derived as  \cite{RefAS}
\begin{align}
\langle L_t\rangle &= \sum_{L\ge0} L Q_t(L)
= C_{z^t} \frac{\Lambda}{  (1-z)(1-\Lambda)  }  \\
&\simeq
 \left\{\begin{array}{ll}
   \frac{\alpha}{\beta-\alpha-\alpha\beta}
    &  \quad (\alpha<\frac{\beta}{1+\beta} ),\\
  2\sqrt{  \frac{\beta t}{\pi (1+\beta )}  }
    &  \quad (\alpha=\frac{\beta}{1+\beta} ),\\
  (\alpha-\beta + \alpha \beta)  t 
  &  \quad (\alpha>\frac{\beta}{1+\beta} ) ,
 \end{array}\right.
\label{eq:Lsim}
\end{align}
for $t\to\infty$.

Inserting Eqs.~\eqref{eq:Y=wxxxv} and \eqref{eq:Q=} into
Eqn.~\eqref{eq:P=QY}, we obtain the
{\it matrix product dynamical state}
\begin{align}
  P_t(\emptyset) &= C_{z^t} \frac{1-\Lambda}{1-z},
\\
  P_t(\tau_L\cdots\tau_1) 
  &=  C_{z^t} 
  \frac{1-\Lambda}{1-z}  \Lambda^L
  \langle W| X_{\tau_L} \cdots X_{\tau_1}  |V\rangle .
\end{align}
When $\alpha<\frac{\beta}{1+\beta}$ (convergent phase), the matrix
product dynamical state converges to the matrix product stationary
state \cite{RefAY}
\begin{align}
  & \lim_{t\to\infty} P_t(\emptyset)
  = \lim_{z\to1}(1-\Lambda)
  = \frac{\beta-\alpha-\alpha\beta}{\beta(1-\alpha)},
\\
\nonumber
  & \lim_{t\to\infty} P_t(\tau_L\cdots\tau_1)
  = \lim_{z\to1}(1-\Lambda) \Lambda^L
  \langle W| X_{\tau_L} \cdots X_{\tau_1}  |V\rangle \\
  &=  \frac{\beta-\alpha-\alpha\beta}{\beta(1-\alpha)}
  \left[\frac{\alpha}{(1-\alpha)\beta}\right]^L
  \langle W| X_{\tau_L} \cdots X_{\tau_1}  |V\rangle .
\end{align}


\section{Waiting time}
\label{sec:waiting}

The waiting time is one of the most important quantities in queueing theory,
which corresponds to the number of time steps
that a particle needs to leave the system
after entering the system.

Before we derive the waiting time distribution we determine
the distribution of the number $N$ of particles in the system.
In standard queueing theory $N$ is always identical to the length $L$
of the system since the queue has no internal structure. In the EQP
we only know that, by definition, $N$ can not be larger than $L$.
The probability that the number of particles is $N=0$ at
time $t$ is, of course, equal to $Q_t(0)$.
For $N\in \mathbb N$, we find 
\begin{widetext}
\begin{align}
\nonumber
  P_t^{A+B} (N) &= P_t^{A} (N) +P_t^{B} (N) =
 \sum_{ \tau_L\cdots \tau_1 \in\widetilde{S}:\atop
 \#\{j|\tau_j=1\} = N } P_t (\tau_L\cdots\tau_1)
= C_{z^t}  \frac{1-\Lambda}{1-z}
  \bra   (\Lambda D+\Lambda^2 DE)^N  \ket  \\
\label{eq:ProbN=}
&=  C_{z^t} \frac{\Lambda(1-\Lambda)(1+\beta\Lambda)}
{(1-z)} [\Lambda(1-\beta+\beta\Lambda)]^{N-1}, 
\end{align}
where $P_t^{A} (N)$ [resp. $P_t^{B}(N)$] is the probability of finding 
$N$ particles in the system
and the site 1 being occupied (resp. empty) at time $t$.
We calculate $P_t^{A} (N)$ and  $P_t^{B}(N)$ as well:
\begin{align}
\label{eq:ProbAN=}
   P_t^A(N \in \mathbb N) 
 &=  C_{z^t} \frac{1-\Lambda}{1-z}
 \bra   (\Lambda D+\Lambda^2 DE)^{N-1} \Lambda D\ket
  = C_{z^t}   \frac{ \Lambda(1-\Lambda)}{(1-z)}
  [\Lambda(1-\beta+\beta\Lambda)]^{N-1} , 
\\ 
\label{eq:ProbBN=}
  P_t^B(N \in \mathbb N ) 
 &=   P_t^{A+B} (N) - P_t^A(N)
  = C_{z^t} \frac{ \beta\Lambda^2 (1-\Lambda) }{ 1-z }
  [\Lambda(1-\beta+\beta\Lambda)]^{N-1}  ,\\
  P_t^B(0) &=  Q_t(0)  = C_{z^t}\frac{1-\Lambda}{1-z}.
\label{eq:ProbB0=}
\end{align}
\end{widetext}
We also set $P_t^A(0)=0$. Indeed Eqs.~\eqref{eq:ProbN=}, \eqref{eq:ProbAN=} 
and \eqref{eq:ProbBN=} agree with the results derived in \cite{RefAS}
in a more complicated way.
By using the result \eqref{eq:ProbN=},
 the average number of particles 
at time $t$ is found to be \cite{RefAS}  
\begin{align}
\begin{split}
 \langle N_t\rangle 
& =\sum_{N\ge1} N P^{A+B}_t(N) \\
& = C_{z^t} \frac{\Lambda}{(1-z)(1-\Lambda)(1+\beta\Lambda)}
\end{split}
\\
& \simeq  
 \left\{\begin{array}{ll}
   \frac{\alpha(1-\alpha)}{\beta-\alpha-\alpha\beta} 
   & \ \  (\alpha<\frac{\beta}{1+\beta}  ),\\
   2\sqrt{ \frac{\beta t}{\pi(1+\beta)^3}  } 
     & \ \  (\alpha=\frac{\beta}{1+\beta} ),\\
   \frac{\alpha-\beta +\alpha\beta}{1+\beta} t
    & \ \ (\alpha>\frac{\beta}{1+\beta} ) ,
 \end{array}\right.
\end{align}
for $t\to\infty$.

Now we turn to 
the waiting time, i.e.\ the time that a
new particle stays in the system.  For a given number $N$ of particles
in the system, the probability that the waiting time is $T\in\mathbb
N$ is given by
\begin{align}
\label{eq:ATN}
& (A):\ \binom{T-N}{N} \beta^{N+1} (1-\beta)^{T-2N} 
 =\mathcal A(T,N) , \\
& (B):\  \binom{T-N-1}{N}\beta^{N+1} (1-\beta)^{T-2N-1} 
 =\mathcal B(T,N) .
\label{eq:BTN}
\end{align}
Here $A$ (resp. $B$) corresponds to the case
where the rightmost site is
occupied (resp. empty).
Note that $\binom{x}{y} $ denotes the
binomial coefficient, which should not be confused with a
two-dimensional column vector.
The average waiting times
\emph{for given $N$} in the cases $A$ and
$B$ are, respectively,
\begin{align}
\begin{split} 
& \langle T_{N,A} \rangle  
= \sum_{T\ge 2N} T \mathcal A(T,N)
= \frac{N+1}{\beta}  + N -1,
\end{split} \\
\begin{split}
& \langle T_{N,B} \rangle
= \sum_{T\ge 2N+1} T \mathcal B(T,N)
= \frac{N+1}{\beta}  + N  .
\end{split}
\end{align}
This result can be interpreted as follows:
\begin{itemize}
\item[$\bullet$] For each particle,
 it takes one time step to move from
  site 2 to site 1.  For $N+1$ particles, it takes,
  in total, $N$ time
  steps (resp. $N+1$ time steps)
   for the case $A$ (resp. $B$).
\item[$\bullet$] For each particle, it takes $\frac{1}{\beta}$ time
  steps in average to leave the system after arriving at site 1.  For
  $N+1$ particles, it takes, in total, 
   $\frac{N+1}{\beta}$ time steps.
\item[$\bullet$] A new particle entering the system at time $t$ does
  {\it not} wait during time $t$ and $t+1$.  Thus we have to subtract
  1 from the above.
\end{itemize}

Let us consider the probability $W_t (T)$ of the waiting time $T$ for a particle entering the system at time $t$.
Using Eqs.~\eqref{eq:ProbAN=}, \eqref{eq:ProbBN=}, 
 \eqref{eq:ProbB0=},
 \eqref{eq:ATN} and \eqref{eq:BTN},
we find
\begin{align}
\label{eq:W=}
\nonumber
 W_t (T)  &=
 \sum_{N=0}^{ \lfloor T/2 \rfloor}
   \left[  \mathcal A (T,N) P_t^A(N)
     +\mathcal B (T,N) P_t^B(N) \right] \\
&=  C_{z^t} 
    \frac{\beta(1-\Lambda) }{1-z}
    (  1- \beta + \beta \Lambda )^{T-1},
\end{align}
where $\lfloor\cdot\rfloor$ denotes the floor function, i.e.\
$\lfloor T/2 \rfloor = T/2$ (if $T\in 2\mathbb N$)
or $\lfloor T/2 \rfloor = (T-1)/2$
 (if $T\in 2\mathbb N-1$).
In the convergent phase,
$  W_t (T) $ converges to 
 the stationary distribution of the waiting time
\begin{align}
\begin{split}
\lim_{t\to\infty} W_t (T)
&= \lim_{z\to 1}\beta(1-\Lambda)
  \left( 1-\beta+\beta\Lambda  \right)^{T-1}   \\
&= \frac{\beta-\alpha-\alpha\beta}{1-\beta+\alpha\beta}
   \left( \frac{1-\beta+\alpha\beta}{1-\alpha}\right)^T,
\end{split}
\end{align}
which agrees with the result in \cite{RefY}.

To finish this section, we investigate the average waiting time:
\begin{align}
 \langle T_t \rangle &= 
C_{z^t}   \frac{\beta(1-\Lambda) }{1-z}
 \sum_{T\ge1}T (  1- \beta + \beta \Lambda )^{T-1} \nonumber\\
    &= C_{z^t} \frac{1}{\beta(1-z)(1-\Lambda)} .
\end{align}
The order of the closest singularity $z=1$ to the origin 
depends on the parameters $(\alpha,\beta)$ \cite{RefAS}:
\begin{align}
  \lim_{z\to 1} \frac{1-z}{\beta(1-z)(1-\Lambda)} &=
  \frac{1-\alpha}{\beta-\alpha -\alpha \beta}
  \quad ( \alpha < \frac{\beta}{1+\beta} ),\!\!\! \\
  \lim_{z\to 1} \frac{(1-z)^{\frac{3}{2}}  }
  {\beta(1-z)(1-\Lambda)} &=
  \sqrt{  \frac{1}{\beta(1+\beta)}   }
  \quad ( \alpha = \frac{\beta}{1+\beta} ),   \\
  \lim_{z\to 1} \frac{(1-z)^2}{\beta(1-z)(1-\Lambda)} &=
  \frac{\alpha-\beta+\alpha\beta} {\beta}
  \quad ( \alpha > \frac{\beta}{1+\beta} ), \!\!\! 
\end{align}
and thus we have
\begin{align}
   \langle T_t \rangle &\to
      \frac{1-\alpha}{\beta-\alpha -\alpha \beta}
  \qquad ( \alpha < \frac{\beta}{1+\beta} ), \\
   \langle T_t \rangle &=
   2  \sqrt{  \frac{t}{\pi\beta(1+\beta)}   }  + o(\sqrt{t})
  \qquad ( \alpha = \frac{\beta}{1+\beta} ),  \\
   \langle T_t \rangle &=
     \frac{\alpha-\beta+\alpha\beta} {\beta}t + o(t)
  \qquad ( \alpha > \frac{\beta}{1+\beta} ), 
\end{align}
as $t\to \infty$.
We note that
one of the central results of queueing theory,
Little's theorem \cite{RefMedhi}, is indeed satisfied
in the convergent phase ($\alpha<\frac{\beta}{1+\beta}$) \cite{RefY}:
\begin{align}
 \alpha \lim_{t\to \infty} \langle T_t \rangle
 =   \lim_{t\to\infty} \langle N_t\rangle . 
\end{align}
We also notice that, in the divergent phase and on the critical line
($\alpha\ge\frac{\beta}{1+\beta}$) as well as in the convergent phase,
there is a physically natural relation between the average waiting
time and the average number of particles:
\begin{align}\label{eq:JT=N}
   J_{1 t}\cdot \langle T_t \rangle
   \simeq  \langle N_t \rangle \qquad 
   (t\to\infty).
\end{align}
Here $ J_{1 t} $ is the current of particles passing through the exit (outflow),
which will be derived in the next section.
Note that this relation
also holds for the usual queueing process, see Appendix.


\section{Density and current}\label{sec:density-current}

We consider the probability $\rho_{jt}$ that the site $j$ is occupied
at time $t$, i.e.\ the density profile.  The initial condition implies
that $ \rho_{jt} =0$ for $j>t$ and $\rho_{t t}=\alpha^t$.  The density
profile for general $j$ and $t$ can be calculated as
\begin{widetext}
\begin{align} \label{eq:den-pro}
\begin{split}
&   \rho_{j t}
  = \sum_{\tau_k=0, 1}P_t(1\tau_{j-1}\cdots\tau_1) 
       +\sum_{\tau_k=0,1 \atop L\ge j+1 }
   P_t(1\tau_{L-1}\cdots \tau_{j+1} 1 \tau_{j-1} \cdots \tau_1) \\
  &=  C_{z^t}  \frac{1-\Lambda}{1-z}  \Big[  \Lambda^j
     \langle W| D (D+E)^{j-1}  |V\rangle  
   +\sum_{L\ge j+1}   \Lambda^L 
   \langle W| D(D+E)^{L-j-1}D(D+E)^{j-1} |V\rangle
  \Big] \\
  &= C_{z^t}  \frac{1-\Lambda}{1-z} \Lambda^j \Big\{
     \langle W| D (D+E)^{j-1}  |V\rangle  
   + \Lambda\langle W| D
    \left[ 1- \Lambda(D+E) \right]^{-1}
        D(D+E)^{j-1} |V\rangle  \Big\}\\
  &=  C_{z^t} \frac{\Lambda^j}{(1-z)(1+\beta\Lambda)}.
\end{split}
\end{align}
\end{widetext}
By definition, the particle current $J_{1 t}$ passing through the exit
(the right end) during $t$ and $t+1$ is given by
\begin{align}
\label{eq:cur-pro}
   J_{1t}  = \beta  \rho_{1t} .
\end{align}
The particle current $J_{jt}$ through the bond between the sites
$j(\ge 2)$ and $j-1$
\begin{align}
\begin{split}
  J_{jt} =&
    \sum_{\tau_k=0, 1}P_t(10\tau_{j-2}\cdots\tau_1)  \\ 
       &+\sum_{\tau_k=0,1 \atop L\ge j+1 }
P_t(1\tau_{L-1}\cdots\tau_{j+1}10\tau_{j-2}\cdots\tau_1) 
\end{split}
\end{align}
also satisfies the  relation
\begin{align}
  J_{jt} = \beta \rho_{jt}
\end{align}
since $ DE(D+E)^{j-2} |V\rangle = \beta D(D+E)^{j-1} |V\rangle $ .

For the generic choice of parameters $\alpha$ and $\beta$, the density
profile near the right end converges as
\begin{align}
  \rho_{jt}\to 
 \lim_{z\to 1}\frac{\Lambda^j}{1+\beta\Lambda} 
 =
  \begin{cases}
    \frac{1}{1+\beta} 
      &  (\alpha \ge \frac{\beta}{1+\beta}), \\
  (1-\alpha)\left[ \frac{\alpha}{(1-\alpha)\beta} \right]^j &
       (\alpha < \frac{\beta}{1+\beta}),
  \end{cases}
\end{align}
for $t\to\infty$.
Here we took the limit with the site number $j$ independent of time $t$.
In particular, for $j=1$, we have
\begin{align}
   \lim_{t\to\infty}  \rho_{1t} =
  \begin{cases}
 \frac{1}{1+\beta}&(\alpha \ge \frac{\beta}{1+\beta}), \\
\frac{\alpha}{\beta}  &  (\alpha < \frac{\beta}{1+\beta}),
  \end{cases}
\\
   \lim_{t\to\infty}  J_{1 t}  =
  \begin{cases}
\frac{\beta}{1+\beta}&(\alpha\ge\frac{\beta}{1+\beta}), \\
\alpha &  (\alpha < \frac{\beta}{1+\beta}) ,
  \end{cases}
\label{eq:limJ1t}
\end{align}
confirming the relation \eqref{eq:JT=N}.

\begin{figure}\begin{center}
\centering
\includegraphics[width=0.9\columnwidth]{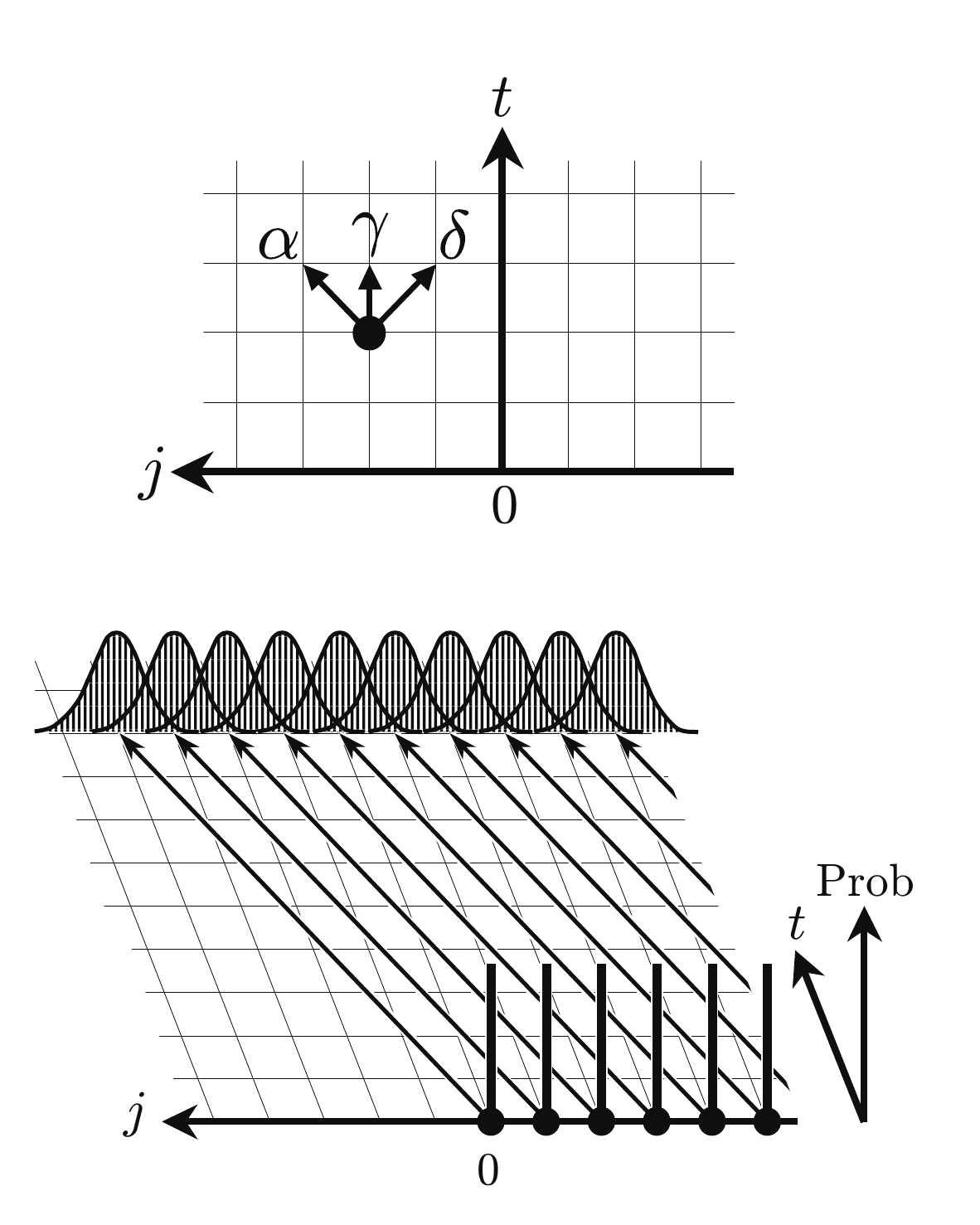}
\caption{Schematic picture of the 
  noninteracting-random-walker interpretation.  }
\label{fig:schematic} 
\end{center}\end{figure}

Let us now consider rescaled density profiles in the divergent phase
and on the critical line, where the average system length grows of
order $t$ and $\sqrt{t}$,
respectively [see Eqn.~\eqref{eq:Lsim}].  
First we observe
 that the density profile $\rho_{jt}$ can be
interpreted as the expected number of noninteracting asymmetric random 
walkers at time $t$ on site $j$
 since  the expression
\eqref{eq:den-pro} satisfies the equation
\begin{align}
  \rho_{j,t+1} =
  \alpha \rho_{j-1,t} +
   \gamma \rho_{jt} + \delta \rho_{j+1,t} 
\end{align}
with $\gamma=(1-\alpha)(1-\beta)$ and $ \delta=(1-\alpha) \beta $,
which is the same as for $Q_t(L)$ [cf.~\eqref{eq:QL}].  
We extend the domain $j\in\mathbb N$  
to $j\in\mathbb Z$ so that $\rho_{jt}$ can be regarded as the expected
number of walkers
with the initial condition 
that a random walker exists at each site $i\in\mathbb Z_{\le 0}$ with
probability
\begin{align}
  \rho_{i0}
  = \frac{1}{1+\beta} + 
  (1-\alpha)\left[\frac{\alpha}{(1-\alpha)\beta}\right]^i
  - \frac{1-\alpha -\alpha\beta }{1+\beta}
     \frac{1}{(-\beta)^i}\,   .
\end{align}
Each walker at site $j$ hops to its left site $j+1$ with probability
$\alpha$, to the right site $j-1$ with probability $\delta$, or stays 
at site $j$ with probability $\gamma$ ($\alpha+\gamma+\delta=1$), see
Fig.~\ref{fig:schematic}.
Let $\epsilon^{(i)}_{jt}$ be the probability that the walker starting
from the site $i$ is in site $j$ at time $t$, which is distributed
around $Vt+i$ as \cite{RefKRB}
\begin{align}\label{eq:eps=}
\epsilon^{(i)}_{jt} \simeq
 \frac{1}{\sqrt{2\pi\sigma t}} 
 \exp \left[ - \frac{ (j-Vt-i)^2 }{2\sigma t} \right]
\end{align} 
for the generic case $0<\alpha<1$ and $0<\beta<1$.
Here $\sigma= \alpha+\delta -(\alpha-\delta)^2$,
and  $V=\alpha-\delta$ 
which is equal to the velocity for the system length,
see Eqn.~\eqref{eq:Lsim}.
The density profile is expressed as
\begin{align}
   \rho_{jt} =
   \sum_{i \le 0  }  \rho_{i0} \epsilon^{(i)}_{jt}.
\end{align}

\begin{figure}\begin{center}
\centering
\includegraphics[width=1\columnwidth]{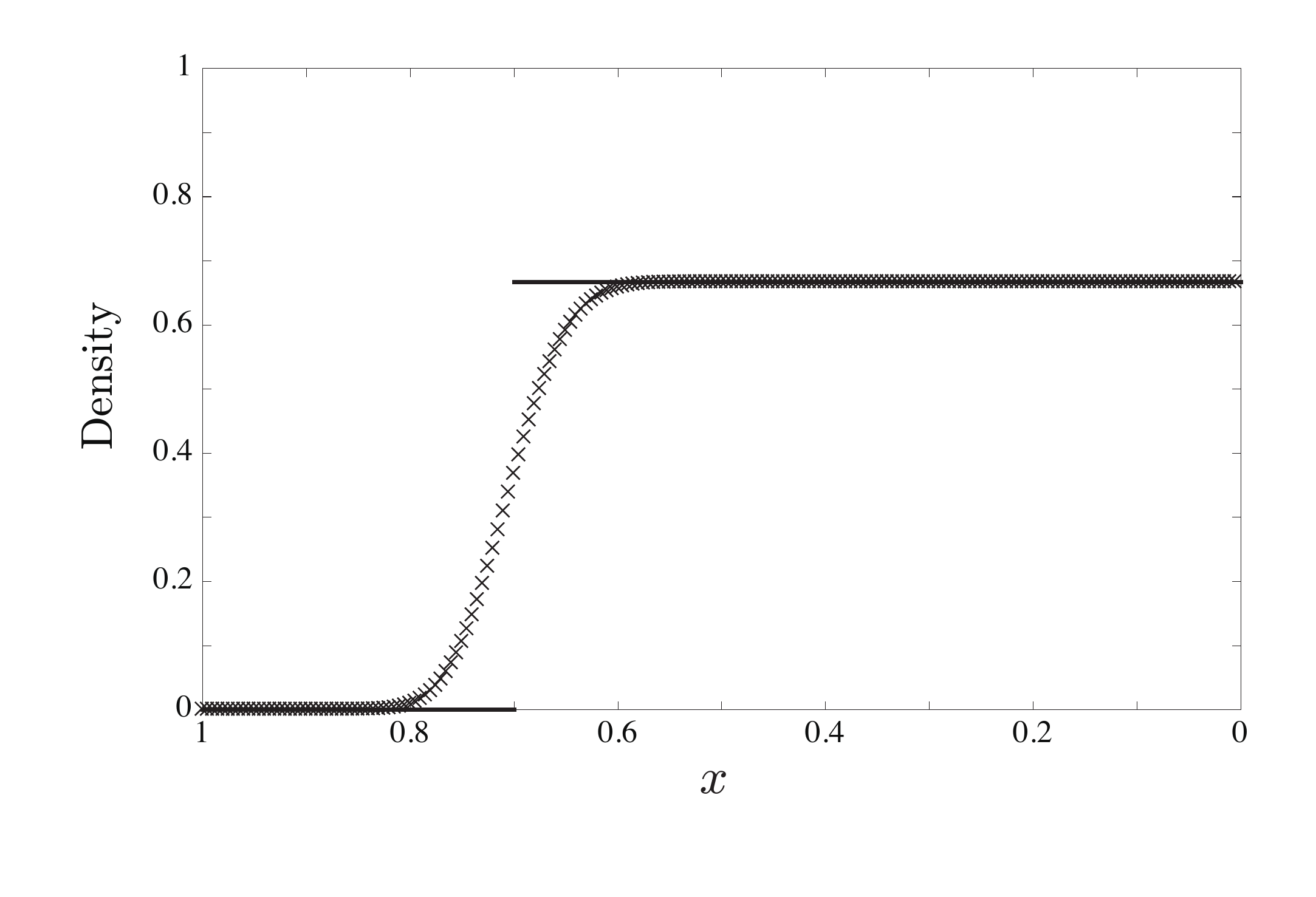}
\vspace{-12mm}
\caption{Rescaled density profile in the divergent phase.
  The parameters are chosen as $(\alpha,\beta)=(4/5,1/2)$.  The
  markers $\times$ correspond to Eqn.~\eqref{eq:den-pro} with
  $t=200,x=j/t$, and the line to the asymptotic form \eqref{asym-div}.
  }
\label{fig:den-div} 
\end{center}\end{figure}

In the divergent phase
$\alpha>\frac{\beta}{1+\beta}$ (with
$\alpha<1$ and $0<\beta<1$),
noting the initial condition
\begin{equation}
\lim_{i\to -\infty} \rho_{i0} =\frac{1}{1+\beta} 
\end{equation}
and the form \eqref{eq:eps=},
we find that the density profile with rescaling
of the position $j=xt$ converges as
\begin{align}\label{asym-div}
   \rho_{xt, t} \to
  \begin{cases}
     \frac{1}{1+\beta} &\ \  (0<x<V) ,  \\
     0 &\ \    (V<x<1) .
  \end{cases}
\end{align}
Figure~\ref{fig:den-div}
gives an example for the rescaled density profile in the divergent phase.

\begin{figure}\begin{center}
\includegraphics[width=1\columnwidth]{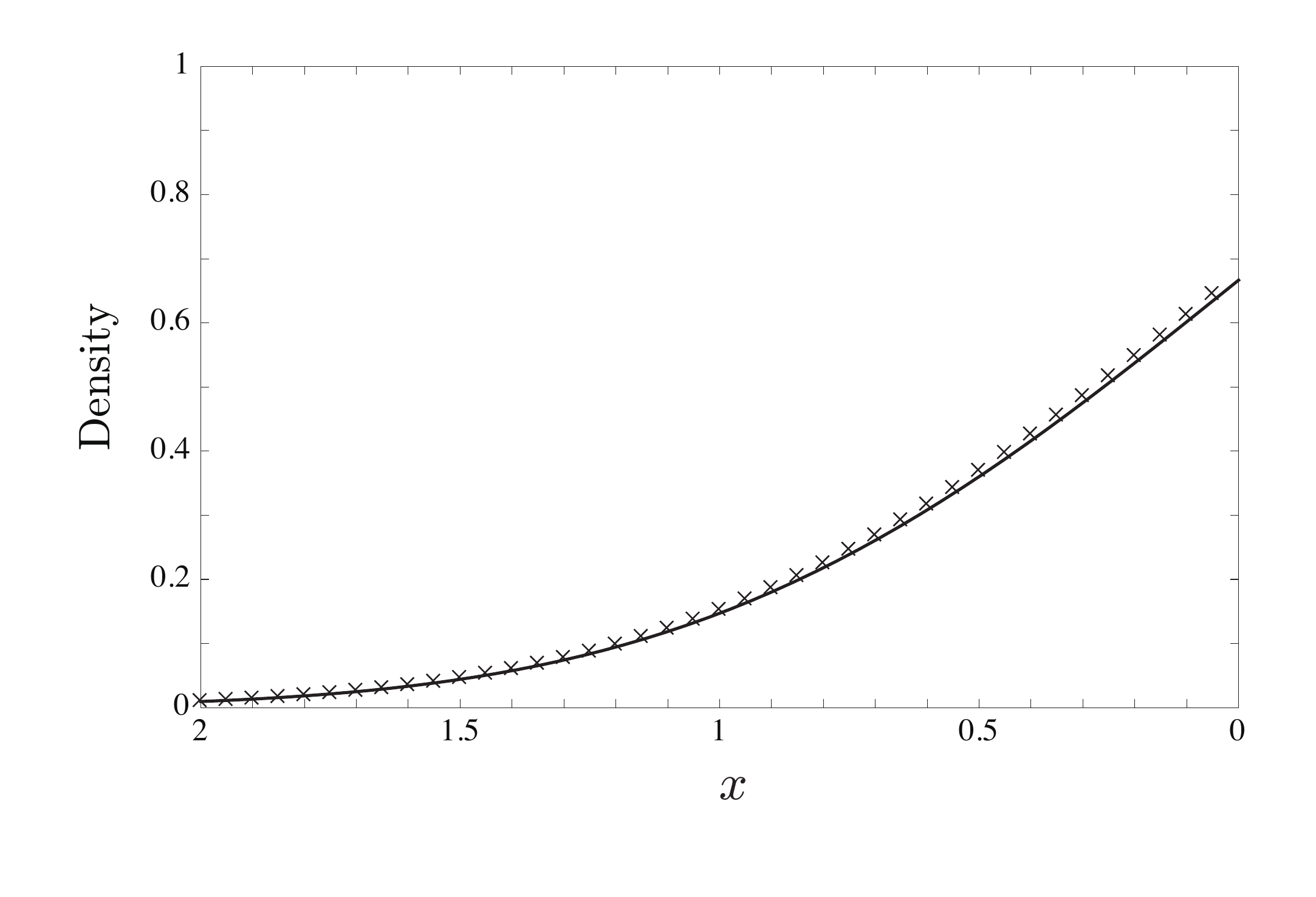}
\vspace{-12mm}
\caption{Rescaled density profile on the critical line.
  The parameters are chosen as $(\alpha,\beta)=(1/3,1/2)$.  The
  markers $\times$ correspond to Eqn. \eqref{eq:den-pro} with
  $t=400,x=j/\sqrt{t}$, and the line to the asymptotic form
  \eqref{asym-crit}.  }
\label{fig:den-crit} 
\end{center}\end{figure}

On the critical line
 $\alpha=\frac{\beta}{1+\beta}$ ($0<\beta< 1$),
noting the initial condition 
\begin{equation}
\lim_{i\to -\infty} \rho_{i0} =\frac{2}{1+\beta}
\end{equation}
and the form \eqref{eq:eps=} with $V=0$,
we find that 
the density profile \eqref{eq:den-pro} with the rescaling
$x=\frac{j}{\sqrt{t} }$ converges as
\begin{align}
\label{asym-crit}
  & \rho_{x\sqrt{t},t}
    \to \frac{1}{1+\beta} \operatorname{erfc}
  \left( \frac{x}{2}\sqrt{\frac{1+\beta}{\beta}  } \right) . \end{align}
Here $\operatorname{erfc}$ is the complementary error function:
$\operatorname{erfc}(x)=\frac{2}{\sqrt{\pi}}\int^{\infty}_{x}e^{-y^2}dy$.
Figure~\ref{fig:den-crit} gives an example for the rescaled density
profile on the critical line.

Now we consider some special cases.
When $\alpha=1$, the position of
the leftmost particle is $t$ by definition, and we have $\rho_{tt}=1$.
In this case, $\Lambda=z$
 and the density profile
 becomes simply  
\begin{align}
\rho_{jt} =& C_{z^t}   \frac{z^j}{(1-z)(1+\beta z)} 
  =  \frac{1-(-\beta)^{t-j+1}}{1+\beta}.
\end{align}
In particular, the density profile observed by the leftmost particle
is independent of time $t$, and exhibits oscillations.

\begin{figure}\begin{center}
\includegraphics[width=1\columnwidth]{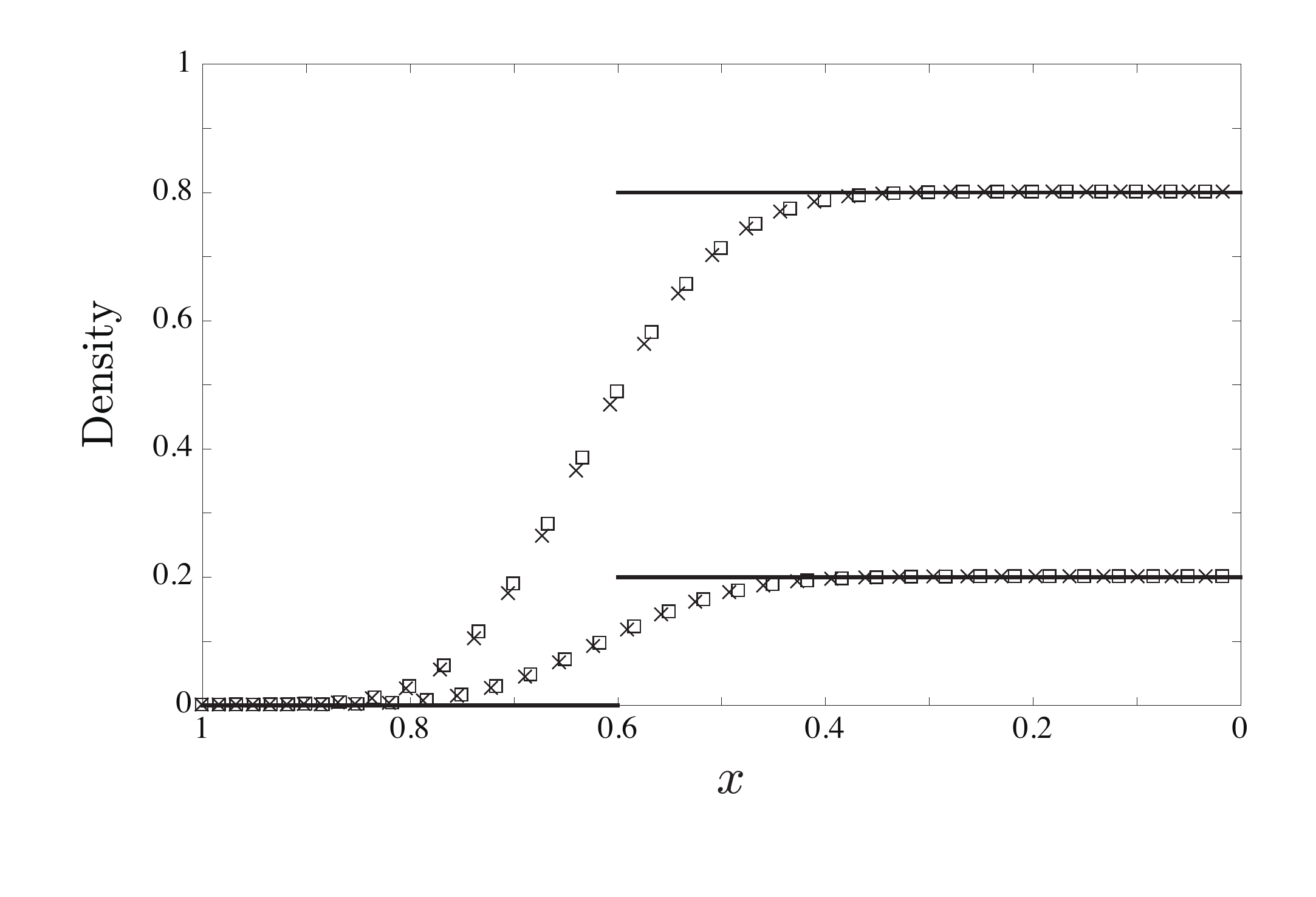}
\vspace{-12mm}
\caption{Rescaled density profile
  for $(\alpha,\beta)=(4/5,1)$.
The markers $\square$ and $\times$
  correspond to Eqn.~\eqref{eq:den-pro}
   with $t=60$ and $61$,
  respectively, and $x=j/t$.
The line corresponds to the 
asymptotic form \eqref{asym-b=1}.  }
\label{fig:b=1} 
\end{center}\end{figure}

When $\beta=1$ and $\alpha>\frac{1}{2}$,
another oscillation occurs.
Since $\gamma=0$,
the walker starting from the site $i$
can exist on the site $j$ at time $t$ only
if $j-i-t\in 2\mathbb Z$,
and is distributed around $Vt+i$ at time $t$ as
\begin{align}\label{eq:eps2}
\epsilon^{(i)}_{jt} \simeq
 \sqrt{\frac{2}{\pi\sigma t}} 
 \exp \left[ - \frac{ (j-Vt-i)^2 }{2\sigma t} \right] .
\end{align} 
We also note that the initial condition
for the walker on site $i$ converges as 
\begin{align}
 \lim_{i\to -\infty\atop i\in 2\mathbb Z+1}\rho_{i0}
  = 1-\alpha , 
 \lim_{i\to -\infty\atop i\in 2\mathbb Z}\rho_{i0}
  = \alpha .
\end{align}
In the limit $t\to\infty$ with the scaling $j=xt$, we have
\begin{eqnarray}\label{asym-b=1}
\rho_{j t}\to
  \begin{cases}
    1-\alpha
  & (0<x<2\alpha-1,j-t\in 2\mathbb Z +1), \\
    \alpha
  & (0<x<2\alpha-1,j-t\in 2\mathbb Z), \\
    0     &    (2\alpha-1<x<1).
  \end{cases}
\end{eqnarray}
Figure~\ref{fig:b=1}
gives an example for the rescaled density profile
in this case.
Also when $\alpha=\frac{1}{2}$ and  $\beta=1$,
the walker starting from the site $i$
can exist on the site $j$ at time $t$ only
if $j-i-t\in 2\mathbb Z$,
and is distributed around $Vt+i$ at time $t$ as
Eqn.~\eqref{eq:eps2}.
Noting $V=0$ and the initial condition 
$\rho_{i0} =1$,
we find the rescaled density profile
\eqref{asym-crit}.

For the very special case $\alpha=\beta=1$, which is completely
deterministic, the site $j$ is occupied if $j\le t$ and $t-j$ is even,
or empty otherwise.
The density profile oscillates between 1 and 0.

\section{Concluding remarks}\label{sec:conclusion}

We have studied dynamical properties of the EQP with deterministic bulk
hopping $p=1$.
We found the exact dynamical state
 in matrix product
(MP) form with a two-dimensional representation of the matrices and 
vectors.  The MP dynamical state approaches the
MP stationary state in the convergent phase
$\alpha<\frac{\beta}{1+\beta}$ as $t\to\infty$.  We have obtained the
time-dependent distributions of the system length $L$
\eqref{eq:Q=}, the number of particles $N$ \eqref{eq:ProbN=} and the
waiting time $T$ \eqref{eq:W=}, and the time-dependent
density~\eqref{eq:den-pro} and current~\eqref{eq:cur-pro} profiles of
site $j$.  
An interesting point is that essentially they are given in the form
\begin{align}
   C_{z^t}  \Psi(z) \Phi(z) ^x \quad
   (x=L,N,T,j) 
\end{align}
with functions $\Psi$ and $\Phi$ including the square root $r$
[cf.~\eqref{eq:r=}].

We found that the asymptotic density profile in the divergent phase
(with the generic choice of parameters) is flat.  In contrast, the
density profile for $p<1$ is nontrivial \cite{RefASfuture}.  One of
the important tasks for future studies
is to determine the form
of the density profile and its dependence on the system parameters
for the general case.  The stationary state for the EQP with 
probabilistic hopping $p<1$ has a matrix product form with infinite
dimensional matrices \cite{RefAY}.  This fact makes us expect that the
MP dynamical state can be extended to the $p<1$ case, which approaches
the MP stationary state in the limit $t\to\infty$ as in the following
diagram: \newcommand{\dst}{\displaystyle}
\begin{eqnarray*} \label{DiagComm}
\begin{CD}
\begin{array}{l}
  {\rm dynamical\ state} \\
  {\rm of\ unknown\ form} 
 \end{array} 
 @> \dst p\to 1 >> 
\begin{array}{l}
  {\rm MP\ dynamical\ state} \\
  {\rm with}\  2{\rm D\ matrices}
 \end{array} 
     \\
  @V  \dst  t\to\infty    VV  @VV \dst  t\to\infty  V  \\
\begin{array}{l}
  {\rm MP\ stationary\ state} \\
  {\rm with}\  \infty{\rm D\ matrices}
 \end{array} 
   @> \dst p\to 1 >> 
\begin{array}{l}
  {\rm MP\ stationary\ state} \\
  {\rm with}\  2{\rm D\ matrices}.
 \end{array} 
\end{CD}
\end{eqnarray*}
In the probabilistic hopping case $p<1$, however, the master equation
cannot be simplified similar to Eqs.~\eqref{master-u10}-\eqref{master-u11},
and the factorization ansatz \eqref{eq:P=QY} is no longer valid
\cite{RefASfuture}.

What we have investigated here is a very basic model of queues with
excluded-volume effect.
Apart from the generalization to $p<1$ 
one can consider various other generalizations of
the EQP, for example multi-lane queues with some
types of queue-changing rules.

\begin{acknowledgments}
  C. Arita is a JSPS Fellow for Research Abroad.  The authors thank
  Martin R. Evans and Gunter M. Sch\"utz for useful discussions and
  the Max-Planck-Institut f\"ur Physik komplexer Systeme in Dresden,
  where part of this work was performed, for its hospitality.
\end{acknowledgments}


\appendix

\section{Results on the usual queueing process}

The usual queueing process is characterized by the number $N$ of
particles which is equal to the length of the system since its spatial
structure is not taken into account.  We denote the probability that
the number of particles is $N$ at time $t$ by $P_t(N)$.
At each time,
a particle enters the system with probability $\alpha$.  When a
particle leaves the system at time $t-1$, the next particle can leave
the system at time $t$ with probability $\beta$.  If a new particle
enters the system at time $t$ and there is
no other particle, this
new particle can leave the system simultaneously.  The critical line
for the usual queueing process is $\alpha=\beta$, and the system is
convergent or divergent if $\alpha<\beta$ or $\alpha>\beta$,
respectively.  The dynamical solution to the master equation
\begin{align}
 P_{t+1} (0)   =&\,  
  [  (1-\alpha)  + \alpha\beta] P_t(0)
  + (1-\alpha)\beta P_t(1) , \\
\begin{split}
 P_{t+1}  (N)
  =&\,  (1-\alpha)\beta P_t (N+1)  \\
   &  +~[(1-\alpha)(1-\beta) + \alpha\beta] P_t(N) \\
  &  +~\alpha(1-\beta) P_t (N-1) 
      \qquad (N\in \mathbb N)
\end{split}
\end{align}
with the initial condition 
\begin{align}
   P_0(0)=1,\quad P_0(N) =0 \   (N \in\mathbb N)
\end{align}
is given  by
\begin{align}
  P_t(N)  =  C_{z^t}  \frac{1-\Theta}{1-z} \Theta^N ,
\end{align}
where
\begin{align}
  \Theta =& 
  \frac{1 - (1-\alpha -\beta +2 \alpha\beta )z - s}
  {2(1-\alpha)\beta z}, \\
  s  =&  \sqrt{ [1 - (1-\alpha -\beta +2\alpha\beta )z ]^2
  -4\alpha\beta(1-\alpha)(1-\beta)z^2} .
\end{align}
When $\alpha<\beta $, the system approaches the stationary state
\begin{eqnarray}
  \lim_{t\to\infty} P_t(N) 
   &=&\lim_{z\to 1}  (1-\Theta) \Theta^N \nonumber\\
   &=&  \frac{  \beta(1-\alpha)  }{\beta-\alpha}
   \left[ \frac{\alpha(1-\beta)}{\beta(1-\alpha)} \right]^N.
\end{eqnarray}
The average number of particles at time $t$ is 
\begin{align}
\begin{split}
  \langle N_t \rangle &=  \sum_{N\ge 1}  N P_t(N) 
  = C_{z^t} \frac{\Theta}{(1-z)(1-\Theta)} \\
  &\simeq 
 \left\{\begin{array}{ll}
   \frac{\alpha(1-\beta)}{\beta-\alpha} 
   & \ \  (\alpha<\beta ),\\
   2\sqrt{\frac{\alpha(1-\alpha)t}{\pi}} 
     &  \ \ (\alpha=\beta ),\\
   (\alpha-\beta) t &  \ \ (\alpha>\beta ) .
 \end{array}\right.  
\end{split}
\end{align}
The particle current leaving the system (outflow)
$J_t$  at time $t$ is given by
\begin{align}
\nonumber
 J_t &= \alpha\beta P_t(0) + \beta\sum_{N\ge 1} P_t(N) 
  =  C_{z^t} \frac{ \beta[\Theta+\alpha(1-\Theta)] }{ 1-z }\\
 & \to
  \left\{\begin{array}{ll}
     \alpha & (\alpha< \beta),  \\
    \beta & (\alpha\ge\beta),
  \end{array}\right.
\end{align}
for $t\to\infty$.
Note that, in the EQP case, $ \lim_{t\to \infty} J_{1t} $
\eqref{eq:limJ1t} is not equal to the exit probability $\beta $ in the
divergent phase since the rightmost site can be empty.  The
probability of the waiting time $T$ for a given number $N$ of
particles is $\binom{T}{N} \beta^{N+1} (1-\beta)^{T-N}$
from which we find the
probability of the waiting time for a given time $t$:
\begin{align}
\begin{split}
&\sum_{N=0}^{T} P_t(N) \binom{T}{N}\beta^{N+1} (1-\beta)^{T-N} \\
 & = C_{z^t} \frac{\beta(1-\Theta)}{(1-z)} 
 (1-\beta+\beta\Theta)^T  .
\end{split}
\end{align}
The average waiting time is then
\begin{align}
 &   \langle T_t \rangle  =
    C_{z^t} \frac{\beta(1-\Theta)}{(1-z)}
    \sum_{T\ge 1}  T(1-\beta+\beta\Theta)^T  \\
 &=  C_{z^t}  \frac{1-\beta+\beta\Theta}
 {\beta(1-z)(1-\Theta)} 
  \simeq 
 \left\{\begin{array}{ll}
   \frac{1-\beta }{\beta -\alpha }  &  (\alpha<\beta ),\\
   2 \sqrt{\frac{(1-\alpha) t}{\pi\alpha}} 
     &  (\alpha=\beta ),\\
    \frac{\alpha-\beta}{\beta} t &  (\alpha>\beta ) ,
 \end{array}\right.
\nonumber
\end{align}
and the relation $ J_t\langle T_t \rangle \simeq\langle N_t \rangle$
($t\to\infty $) holds.



\begin{thebibliography}{99}


\bibitem{RefL} T.M. Liggett,
 {\em Stochastic Interacting Systems:
    Contact, Voter and Exclusion Processes}, Springer, New York (1999)

\bibitem{RefD}
B. Derrida,
J. Stat. Mech. P07023 (2007)


\bibitem{RefS}  
     G.M. Sch\"utz,
{\em Exactly Solvable Models for Many-Body Systems Far from Equilibrium}
 in {\em Phase Transitions and Critical Phenomena vol 19.}, C. Domb and
 J.~L.~Lebowitz Ed.,   Academic Press, San Diego  (2001)


\bibitem{RefBE}
R.A. Blythe and M.R. Evans,
J. Phys. A: Math. Gen. \textbf{40}, R333 (2007)

\bibitem{RefSCN}
A. Schadschneider, D. Chowdhury and K. Nishinari,
{\em Stochastic Transport in Complex Systems: From Molecules to Vehicles},
Elsevier Science, Amsterdam (2010)

\bibitem{RefMedhi} J. Medhi, {\em Stochastic Models in Queueing Theory},
Academic Press, San Diego 2003)

\bibitem{RefSaaty} 
T.L. Saaty, {\em Elements of Queueing Theory With Applications},
Dover Publ. (1961)

\bibitem{Reffactory} 
W.J. Hopp, M.L. Spearman, {\em Factory Physics}, McGraw-Hill, Boston (2008)

\bibitem{RefA}
C. Arita, 
Phys. Rev. E \textbf{80}, 051119 (2009)

\bibitem{RefY}
D. Yanagisawa, A. Tomoeda, R. Jiang and K. Nishinari,
JSIAM Lett. \textbf{2}, 61 (2010)

\bibitem{RefAY}
C. Arita and D. Yanagisawa,
J. Stat. Phys. \textbf{141}, 829 (2010)

\bibitem{RefAS}
C. Arita and A. Schadschneider,
Phys. Rev. E \textbf{83} 051128 (2011)



\bibitem{RefStS1}
R.B. Stinchcombe, G.M. Sch\"utz,
Europhys. Lett. \textbf{29}, 663 (1995)

\bibitem{RefStS2}
R.B. Stinchcombe, G.M. Sch\"utz,
Phys. Rev. Lett. \textbf{75}, 140 (1995)

\bibitem{RefSaWa}
T. Sasamoto, M. Wadati,
J. Phys. Soc. Jpn. \textbf{66}, 2618 (1997)

\bibitem{RefSchuetz}
G.M. Sch\"utz,
Eur. Phys. J. B\textbf{5}, 589 (1998)

\bibitem{RefAAMP}
C. Arita, A. Ayyer, K. Mallick and S. Prolhac,
J. Phys. A: Math. Gen. \textbf{44} 335004 (2011)


\bibitem{RefKRB}
P.L. Krapivsky, S. Redner and E. Ben-Naim,
{\em  A Kinetic View of Statistical Physics},
Cambridge University Press, Cambridge
 (2010)


\bibitem{RefASfuture}
C. Arita and A. Schadschneider,
unpublished




\end{thebibliography}
\end{document}